\documentclass[aip, apl, twocolumn, superscriptaddress, reprint]{revtex4-2}

\usepackage{graphicx}
\usepackage{bm}
\usepackage{dcolumn}
\usepackage{amssymb,amsmath}
\usepackage[colorlinks=true,allcolors=blue]{hyperref}
\usepackage{color}
\usepackage{siunitx}
\usepackage[export]{adjustbox}
\usepackage{lipsum}
\newcommand{\cOut}[1]{}
\newcommand{\figref}[2]{\hyperref[#1]{\ref{#1}(#2)}}
\begin{document}

\title{Anatomy of localized edge modes in laterally coupled waveguides}



\author{Vadym Iurchuk}
\email[Corresponding author's e-mail: ]{v.iurchuk@hzdr.de}
\affiliation{Institute of Ion Beam Physics and Materials Research, Helmholtz-Zentrum Dresden-Rossendorf, 01328 Dresden, Germany}

\author{Sven Stienen}
\affiliation{Institute of Ion Beam Physics and Materials Research, Helmholtz-Zentrum Dresden-Rossendorf, 01328 Dresden, Germany}

\author{J\"urgen Lindner}
\affiliation{Institute of Ion Beam Physics and Materials Research, Helmholtz-Zentrum Dresden-Rossendorf, 01328 Dresden, Germany}

\author{Attila K\'akay}
\affiliation{Institute of Ion Beam Physics and Materials Research, Helmholtz-Zentrum Dresden-Rossendorf, 01328 Dresden, Germany}

\date{\today}

\begin{abstract}
We present a systematic micromagnetic study of standing spin-wave modes in infinitely long Permalloy strips with rectangular cross-section. Using a finite-element dynamic-matrix method, we first calculate the eigenfrequencies and the corresponding eigenvectors (mode profiles), as a function of the in-plane magnetic field applied across the strip. The ferromagnetic resonance spectra is computed from the mode profiles, assuming a homogeneous radio-frequency excitation, equivalently to an experimental ferromagnetic resonance measurement. The investigation of the field-dependent mode profiles enables for the classification of the observed resonances, here focusing mostly on the \textit{true edge mode} localized at the vicinity of strip edges. Furthermore, we study the mode localization in pairs of 50-nm-thick Permalloy strips as a function of the strip width and their lateral separation. For closely spaced strips, the spatial profile of the quasi-uniform mode is substantially modified due to a significant hybridization with the edge-localized standing spin-wave modes of the neighbouring strip. We show that a wide-range-tunability of the localized edge-mode resonances can be achieved with a precise control of the magnetostatic coupling between the strips. Extreme sensitivity of the edge mode frequency on the bias field demonstrates a potential of the edge resonances for field sensing. Furthermore, for narrow strips ($\approx$100~nm in width), due to the reduced number of the allowed confined modes, a field-controllable switching between the resonances localized either in the strip center or at the edges of the strips can be achieved.
\end{abstract}

\maketitle

Spin-wave conduits based on ferromagnetic microstrips constitute elementary building blocks of the magnonic architectures for spin-based information transport and processing~\cite{chumakAdvancesMagneticsRoadmap2022}. Numerous studies revealed a rich dynamical spin-wave mode spectrum hosted in confined ferromagnetic microstructures~\cite{bayerSpinwaveExcitationsFinite2005,Publ-Id:29497/1,zhangTuningEdgelocalizedSpin2019,pileNonstationarySpinWaves2022,iurchukTailoringCrosstalkLocalized2023}. Arrays of such microstructures can act as magnonic crystals with magnetic-field controllable excitation bandwidths~\cite{krawczykReviewProspectsMagnonic2014,gallardoSymmetryLocalizationProperties2018,gallardoDipolarInteractionInduced2018,langerSpinwaveModesTransition2019,rychly-gruszeckaShapingSpinWave2022a}. 
Among numerous resonances, which can be excited in ferromagnetic strips, \textit{edge modes} attract a particular attention due to their localized nature, allowing for spin-wave propagation in extremely narrow channels~\cite{laraInformationProcessingPatterned2017,zhangTuningEdgelocalizedSpin2019,iurchukTailoringCrosstalkLocalized2023}, which is beneficial for spin-wave based logic devices.
Recently, a study of the quasi-uniform non-propagating modes in arrays of ferromagnetic strips with various width and separation between the strips showed a significant impact of the magnetostatic coupling on the spin-wave frequencies ~\cite{rychly-gruszeckaShapingSpinWave2022a}. Furthermore, it was also shown that the edge-localized resonances (magnetization dynamics) can also be tuned in a wide frequency range by changing the distance between the neighboring ferromagnetic strips~\cite{iurchukTailoringCrosstalkLocalized2023}.

In this manuscript, we focus on the behavior of the localized standing spin waves (SSW) in isolated 50-nm-thick Permalloy (Py) strips and strip pairs. We study the edge mode localization as a function of a bias magnetic field for different strip widths and separation distances between the strips. We show that for reduced gaps between the strips, the increased magnetostatic coupling largely modifies the resonance fields/frequencies of the modes excited at the inner edges of the strip pair. Furthermore, we show that the presence of the closely spaced neighboring strips considerably alters the quasi-uniform mode profile, manifesting in a mode localization redistribution and to a significant hybridization with the edge-localized standing spin-wave modes. Finally, we show the mode distribution in the strip pairs with reduced width, allowing for a precise confinement of the excited quasi-uniform and localized resonances within the cross-section of the strip.

\begin{figure*}[t]
\centering
    \includegraphics[width=\textwidth]{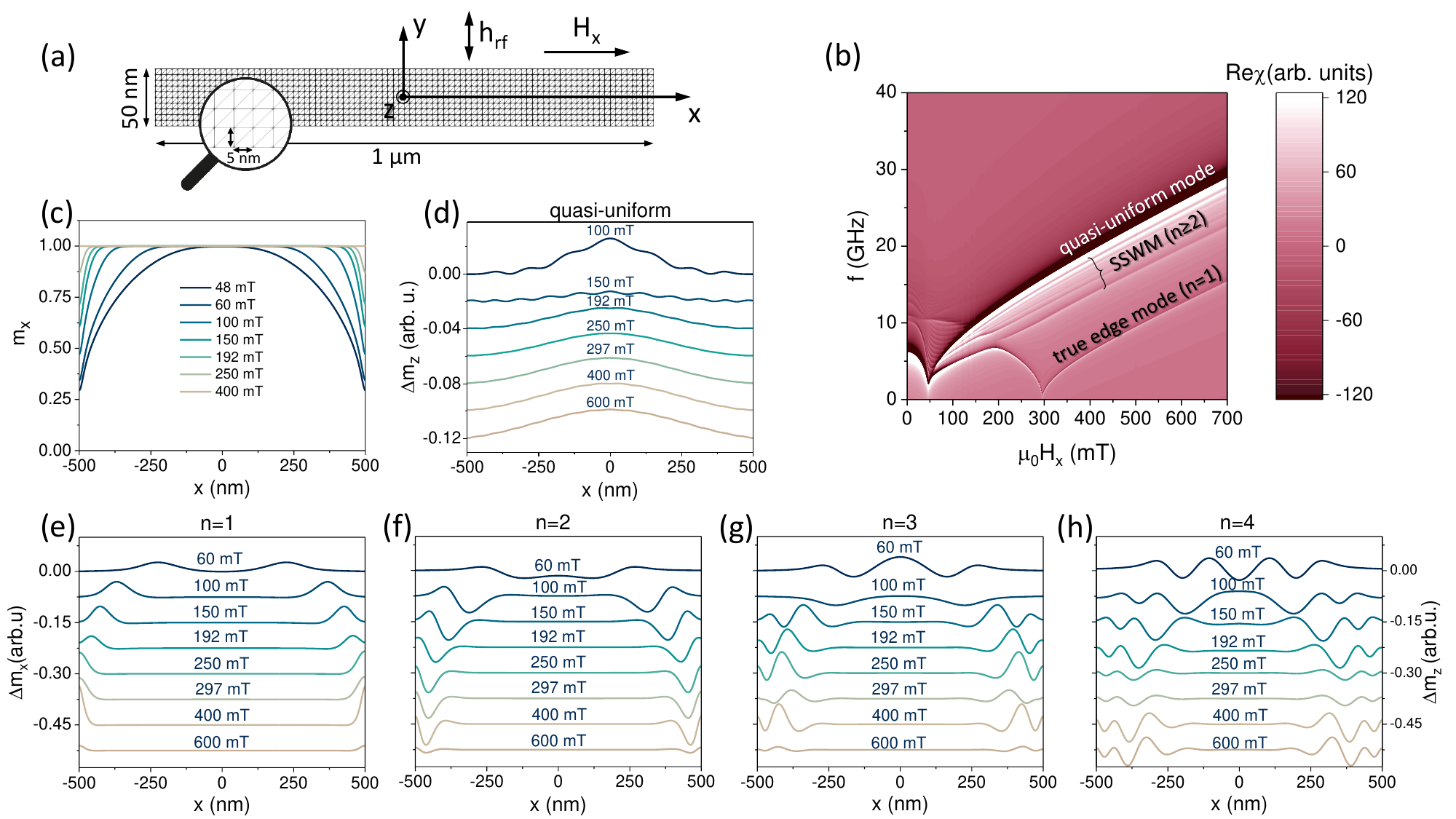}
    \caption{(a) Schematics of the simulation geometry. The bias in-plane field $H_x$ is applied across the strip. The FMR spectra is computed assuming a homogeneous rf field along $Oy$, the out-of-plane direction. (b) Frequency-field FMR absorption map of infinitely long Py strip with \SI{1}{\micro\meter} $\times$ \SI{50}{\nano\meter} cross section. (c) $x$-component of the equilibrium magnetization across the strip width calculated for different $H_x$. (d) Dynamical profiles of the quasi-uniform Kittel mode for the given values of the bias magnetic field $H_x$. Here and further, we plot the dynamical $m_z$ components taken in the middle of the strip cross section ($y$=0). (e--h) Dynamical mode profiles of the first four standing spin-wave modes ($n \leqslant$ 4) for different $H_x$.} 
    \label{fig1}
\end{figure*}

The study is performed with the \textsc{TetraX} finite-element package~\cite{fem_dynmat_SW,tetrax}, that allows to calculate numerically the eigenvalues (the resonance frequencies) and the corresponding eigenvectors (the mode profiles) in infinitely long Py strips with rectangular cross section, among the variety of geometries offered by the program library.
The strip cross-section is discretized into a regular triangular mesh with an average cell size of 5~nm [see Fig.~\figref{fig1}{a}]. We used the following magnetic parameters for Py: saturation magnetization $M_\mathrm{s}$ = 796~kA/m, exchange constant $A_\mathrm{ex}$ = 13~pJ/m, gyromagnetic ratio $\gamma$ = \SI{1.76e11}{\radian/\tesla\second} and Gilbert damping constant $\alpha_\mathrm{G}$ = 0.008. For all simulations, the static magnetic field $H_x$ is applied in-plane along $x$-direction, i.e. across the strip width, while the out-of-plane direction is considered to be the $Oy$ [see Fig.~\figref{fig1}{a}]. For each $H_x$ value, we relax the magnetic state and solve the eigenvalue problem, to get the mode profiles and their frequencies. The mode profiles are used to compute the ferromagnetic resonance (FMR) absorption spectra assuming a homogeneous out-of-plane radio-frequency field $h_\mathrm{rf}$ (for details see \cite{korberSymmetryCurvatureEffects2021}). 


Fig.~\figref{fig1}{b} shows the calculated frequency-sweep FMR absorption map of the infinitely long \SI{1}{\micro\meter} wide and \SI{50}{\nano\meter} thick Py strip as a function of the in-plane magnetic field $H_x$ applied across the strip [see Fig.~\figref{fig1}{a}]. A set of confined spin-wave modes is observed corresponding to the different spin-wave resonances at given magnetic field. The mode with the largest amplitude is the \textit{quasi-uniform mode} or Kittel mode\cite{kittelIntroductionSolidState2005} with a typical hard-axis behavior originating from the presence of the shape anisotropy in the strip confined along the $x$-direction. The lowest-frequency (lowest-energy) mode for a given $H_x$ corresponds to the so-called \textit{true edge mode}~\cite{mcmichaelEdgeSaturationFields2006,Publ-Id:29497/1,iurchukTailoringCrosstalkLocalized2023}, with the magnetization precession localized at the very edge of the strip above saturation [Fig.~\figref{fig1}{e}], as opposed to the higher order edge resonances. The frequency band between the true edge and the quasi-uniform mode is occupied by numerous standing spin-wave modes. For these modes, as seen in Figs.~\figref{fig1}{f--h}, the  spatial magnetization dynamics shifts from the edges towards the strip center, gradually spreading across the strip width with increased mode number.

\begin{figure*}[t]
\centering
    \includegraphics[width=\textwidth]{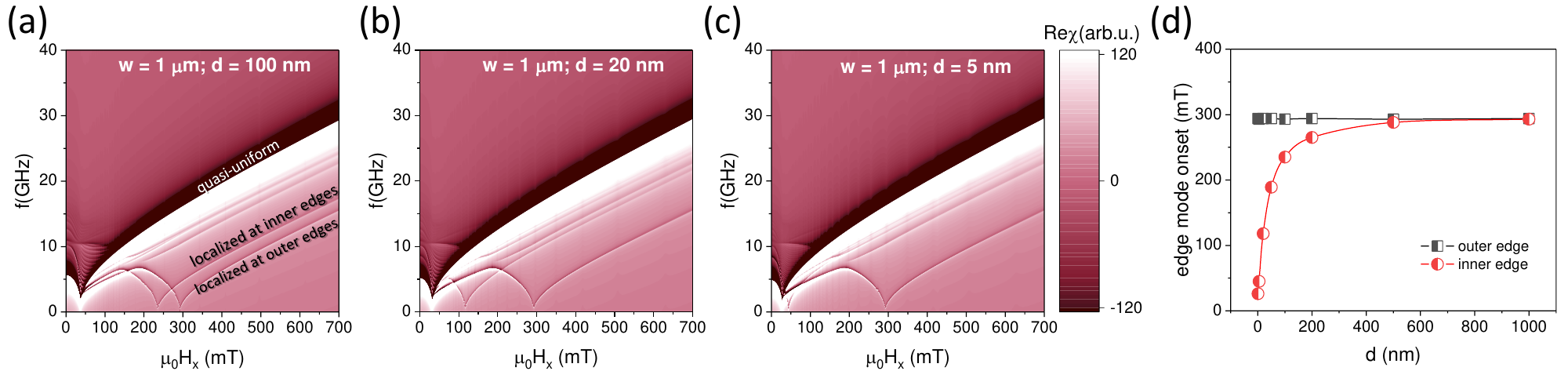}
    \caption{(a--c) Frequency-field FMR absorption maps of the pair of 1-$\mu$m-wide and 50-nm-thick Py strips separated by 100~nm gap (a), 20~nm gap (b) and 5~nm gap (c). (d) Edge saturation fields, corresponding to the onset of the true edge modes localized at the outer (black squares) and inner (red circles) edges of the strip pair as a function of the gap between the strips.}
    \label{fig2}
\end{figure*}

Figs.~\figref{fig1}{e--h} show the mode profiles of the first four observed modes as a function of increased static field. In the absence of a static external field, the waveguide is magnetized at equilibrium in the direction of its long its axis ($Oz$). For already a field of about \SI{48}{\milli\tesla}, corresponding to the \textit{bulk saturation field}, the magnetization in the middle of the strip will rotate into the direction of the field [see Fig.~\figref{fig1}{c}]. In Fig.~\figref{fig1}{b}, this is the point when the lowest mode has its first frequency minimum.
When increasing $H_x$ from \SI{48} to \SI{192}{\milli\tesla}, still following Fig.~\figref{fig1}{b}, the resonance frequency of the lowest mode ($n$=1) increases. During this field increase, as seen in the Fig.~\figref{fig1}{e}, the dynamic regions (or mode amplitudes) continuously shift towards the opposite strip edges, yet the modes become more and more confined in width. The observed gradual confinement together with the pinning of the modes at the edges (where the amplitude is close to zero) leads to an increased exchange interaction contribution to the mode energy. From the other side, in this field region, one can see that the magnetization at the edges continuously rotates in the direction perpendicular to the edges [see Fig.~\figref{fig1}{c}] to reduce the Zeeman energy. However, this results in a continuously increasing demagnetizing field, opposed to the static external magnetic field. 
Apparently, at about \SI{192}{\milli\tesla}, the exchange contribution to the internal fields (from the edge-confinement) reaches it's maximum, but the demagnetizing field will further increase with the static magnetization pointing more and more in the direction of the external field. Moreover, the pinning of the edge modes is gradually released, with increased external field, resulting in a macrospin like dynamics at the edges, from the exchange point of view.
Further increasing the external field in the range of \SI{192}{\milli\tesla} $< H_x <$ \SI{297}{\milli\tesla}, the resonance frequency decreases and shows a well-known dipole-dominated non-aligned edge mode behaviour [see Fig.~\figref{fig1}{c}]. The \textit{edge saturation} eventually occurs at $H_x$ = \SI{297}{\milli\tesla}, the second minimum of the lowest mode, and due to the zero internal fields the frequency goes down to zero, showing a mode softening or representing a so-called Goldstone mode. Above saturation, for $H_x>$ \SI{297}{\milli\tesla}, the $\textit{n}$=1 resonance is the true edge mode, localized in a narrow ($\approx$15--20~nm) region at the strip edges. The true edge mode is also a Kittel-like mode, therefore the further increase of the bias field $H_x$ does not impact the mode localization and leads only to a frequency increase [see Fig.~\figref{fig1}{b}] according to the Kittel (approximately linear for large fields) relation\cite{kittelIntroductionSolidState2005}.
We note, that for the field range 48~mT $< H_x <$ 192~mT, the quasi-uniform mode is hybridized by the higher-order non-aligned modes, which leads to the distinct spatial modulation of its mode profile amplitude [see Fig.~\figref{fig1}{d}].

Higher order standing spin-wave modes ($n \geqslant$ 2) are characterized by an increased number of nodal points and thus the dynamics extends over a larger volume compared to the $n$=1 mode. They exhibit similar behavior, i.e. while nucleated at the strip center as exchange-dominated modes, their intensity maxima gradually propagate towards the strip edges with increased $H_x$ [Figs.~\figref{fig1}{d--f}] as a result of the increased dipolar contribution to the spin-wave energy.

Now that we discussed the main features of the FMR spectra related to the edge modes, we extend our study to a pair of Py strips separated laterally by a gap with a width of $d$. Figs.~\figref{fig2}{a--c} show the frequency-field absorption map of a pair of \SI{1}{\micro\meter} wide and \SI{50}{\nano\meter} thick strips separated by 100, 20 and 5~nm gaps. One can see, that the spectral maps are qualitatively different from that of the single strip. A striking difference is the presence of an additional mode in the low frequency band [see Fig.~\figref{fig2}{a}]. An examination of the mode profiles (not shown here) reveals that this resonance is attributed to a mode, classified as $n$=1i mode, localized at the \textit{inner edges} of the strip pair (see \textcite{iurchukTailoringCrosstalkLocalized2023} for more details).

\begin{figure}[b]
\centering
    \includegraphics[width=0.5\textwidth]{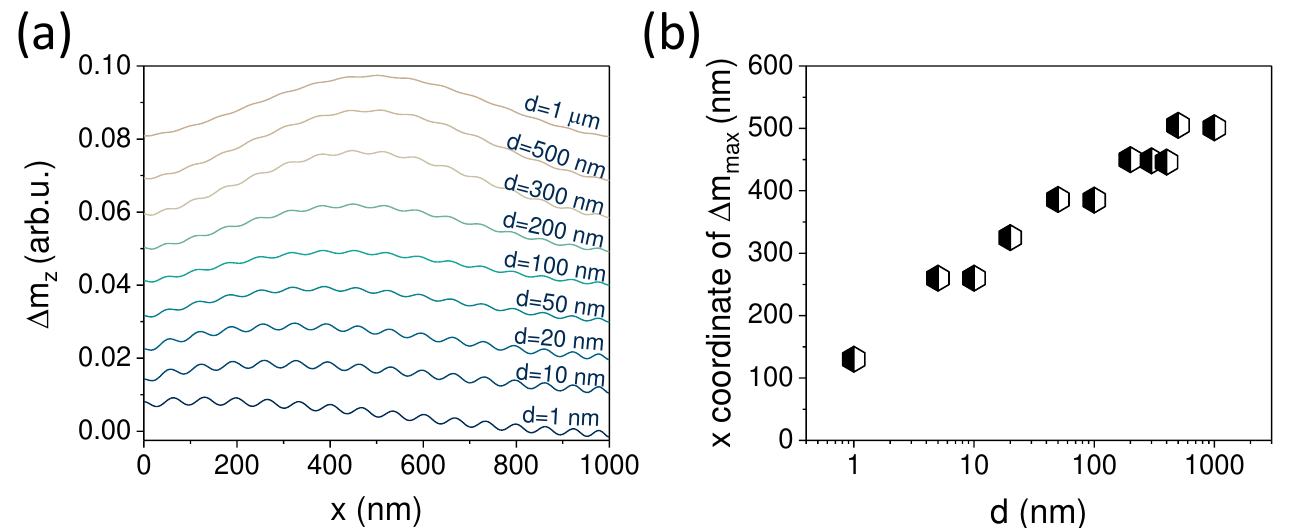}
    \caption{(a) $\Delta m_z$ dynamical component of the quasi-uniform mode for $\mu_0 H_x$=0.6~T as a function of the $x$-coordinate of the right strip for different $d$. The plots are vertically offset for better visibility. (b) Coordinate of the mode intensity maximum versus the gap between the strips.} 
    \label{fig3}
\end{figure}

The $n$=1i mode is also a dipole dominated true edge mode and, therefore, extremely sensitive to the stray field produced by the neighbouring edges. For large gaps ($d \geqslant 1\mu$m), the resonance field/frequency of the $n$=1 and $n$=1i modes are equal, as expected for the completely isolated strips, since the magnetostatic coupling is negligible. As the gap between the strips shrinks, the magnetostatic coupling between the inner edges of the strips increases, favouring the parallel alignment of the inner edges and, thus leading to the reduction of the inner edge saturation field, as summarized in Fig.~\figref{fig2}{d}. The reduction of the resonance field is equivalent to an increase of the resonance frequency. The outer edge saturation field stays unaltered, due to the negligible effect of the stray fields on the opposite edges distant by more than \SI{2}{\micro\meter}s apart. For extremely small gaps $d$ below the exchange length $l_{ex}$ ($l_{ex} \approx$~\SI{5.7}{\nano\meter} for Permalloy), the resonance field of the inner edge mode approaches the shape anisotropy field of the strip pair. For example, at $d$ = 1~nm, the $n$=1i mode can be excited already at $\mu_0 H_x$ = 26~mT, which corresponds to the field needed for the quasi-uniform mode softening. The observed extreme sensitivity of the inner edge mode frequency to the stray fields allows for an efficient use of the edge resonances for field sensing.

\begin{figure*}[t]
\centering
    \includegraphics[width=\textwidth]{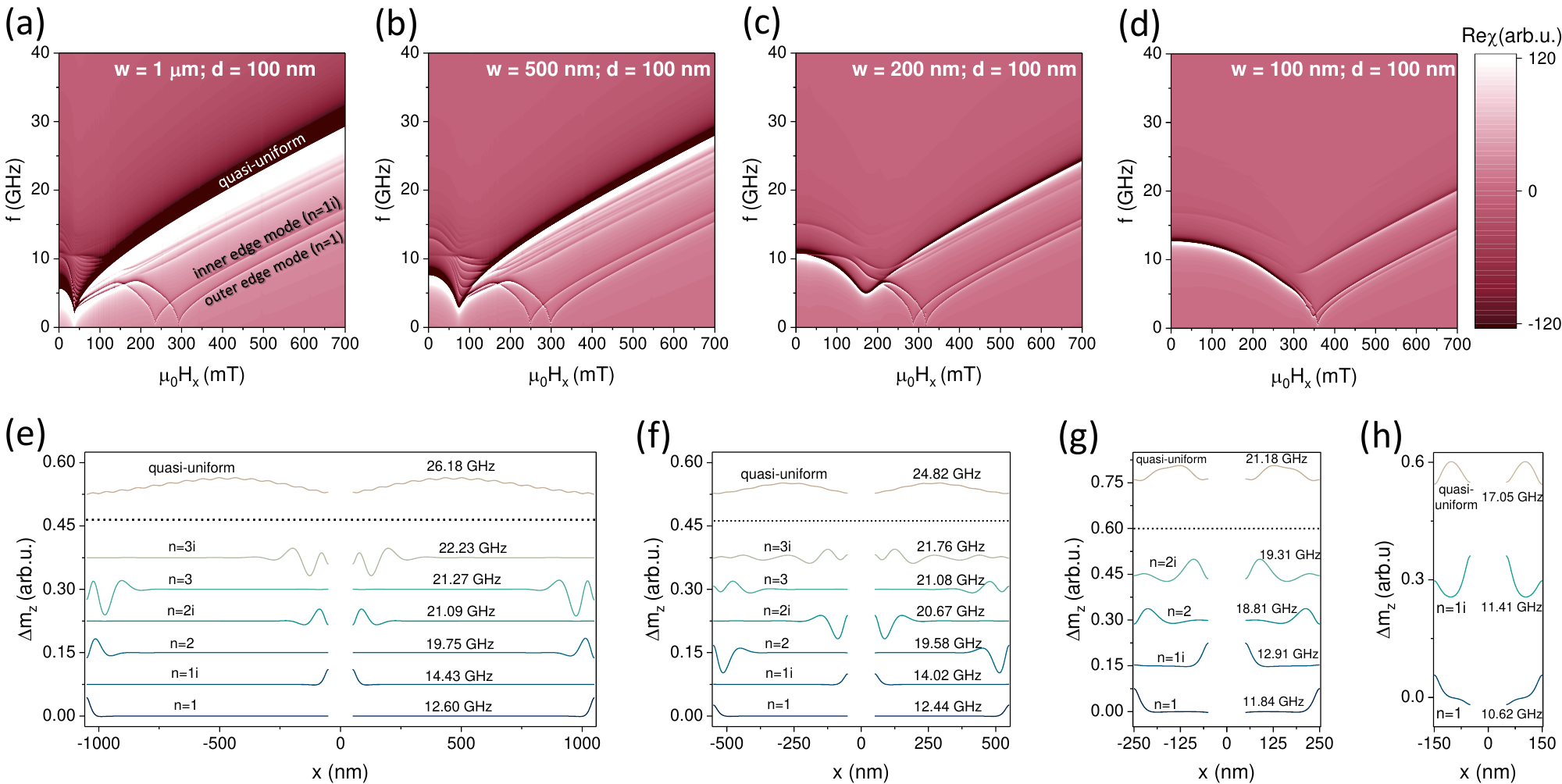}
    \caption{(a--d) Frequency-field FMR absorption maps of the pair of \SI{50}{\nano\meter} thick Py strips with different widths $w$ separated by a \SI{100}{\nano\meter} gap: (a) $w$=\SI{1}{\micro\meter}, (b) $w$=\SI{500}{\nano\meter}, (c) $w$=\SI{200}{\nano\meter}, and (d) $w$=\SI{100}{\nano\meter}. (e--h) Corresponding mode profiles (at $\mu_0 H$ = \SI{0.6}{\tesla}) of the first six localized modes and a quasi-uniform mode for $w$=\SI{1}{\micro\meter} (e) and $w$=\SI{500}{\nano\meter} (f); first four localized modes and a quasi-uniform mode for $w$=\SI{200}{\nano\meter} (g); first two localized modes and a quasi-uniform mode for $w$=\SI{100}{\nano\meter} (h).} 
    \label{fig4}
\end{figure*}

Our analysis shows that not only edge modes are affected by the magnetostatic coupling in the strip pairs, but the intensity and the spatial profile of the quasi-uniform mode is substantially modified too. Fig.~\figref{fig3}{a} shows the quasi-uniform mode profiles versus the strip width for $\mu_\text{0} H_x$=\SI{0.6}{\tesla} and for different gaps $d$ between the strips, ranging from \SI{1}{\nano\meter} to \SI{1}{\micro\meter}. Note, that here we show the spatial mode intensity distribution in the right strip only, whereas the left one shows a completely symmetric picture.
We observe two effects of the gap size on the quasi-uniform mode profiles. Firstly, the gradual shift of the intensity maximum in the direction towards the neighbouring strip, plotted on Fig.~\figref{fig3}{b}. This behavior is in agreement with the results of \textcite{rychly-gruszeckaShapingSpinWave2022a}, where a similar spatial intensity redistribution for the reduced distance between the ferromagnetic strips was reported. It can be explained in terms of the reduced SW pinning at the inner edges of the strips for the increased magnetostatic coupling.

Secondly, we observe a significant modulation of the quasi-uniform mode intensity along the strip width, with a modulation amplitude enhancement for increased magnetostatic coupling [Fig.~\figref{fig3}{a}]. As the gap $d$ decreases, the frequency of the edge modes localized at the inner edges at given $H_x$ increases, gradually approaches the resonance frequency of the quasi-uniform mode [see Figs.~\figref{fig2}{a--c}]. In fact, the frequency of the inner edges will be always higher compared to the frequency of the outer edges for a fixed external field value. Thus the higher order edge modes will be pushed towards the main mode. The accumulation of the higher order modes at the quasi-uniform mode as well as the strong dipole-dipole interaction lead to the hybridization of the quasi-uniform mode with the inner edge-localized modes, which results in the modulation of the mode intensity across the strip width. Note, that the quasi-uniform mode hybridization, due to the presence of the neighbouring strip, is observed even at large bias fields $H_x$ (well above the saturation field). Therefore, this effect is qualitatively different from the hybridization with the higher order modes in the single strip at low bias fields [see Fig.~\figref{fig1}{d}].

Finally, we investigate the mode localization in the strip pairs with a fixed gap $d$ as a function of the strip width $w$. The primary purpose of this investigation is to explore the effect of the width on the localization of the true edge modes and the quasi-uniform mode. On one hand, it is expected that the true edge modes will not anymore localize to the very edge of the sample below a given strip width. On the other hand, the "mode softening" of the quasi-uniform mode is expected to disappear for narrow strips, and the magnetization across the strip width should behave more homogeneously due to the increased proportion of the exchange interaction to the total energy. 

Figs.~\figref{fig4}{a--d} show the frequency-field absorption maps for the pairs of 50~nm-thick Py strips with $w$=\SI{1}{\micro\meter}, \SI{500}{\nano\meter}, \SI{200}{\nano\meter} and \SI{100}{\nano\meter} for $d$ = \SI{100}{\nano\meter}. The corresponding mode profiles of the true edge modes, standing spin-wave modes and the quasi-uniform modes versus the strip width are shown in Figs.~\figref{fig4}{e--h}.
For completeness, we start with the absorption map for the Py strip pair with $w$=\SI{1}{\micro\meter} and $d$ = \SI{100}{\nano\meter} [as also shown in Fig.~\figref{fig2}{a}]. The corresponding mode profiles across the width of the waveguides (the strip pair) for a static external field of \SI{0.6}{\tesla} are extracted and summarized in Fig.~\figref{fig4}{e} for the first six localized modes. The lowest mode in frequency is the true outer edge mode $n$=1, the second lowest is the true inner edge mode $n$=1i, followed by the higher order modes localized close to the edge and, finally, the quasi-static mode. As the strip width is reduced, we observe a gradual decrease of the frequency gap between the $n$=1 and $n$=1i modes for a given field, due to the decreased effective coupling between the inner edges, since the inner-outer edge interaction becomes more significant. Furthermore, when looking on the mode profiles summarized in Fig.~\figref{fig4}{e--h}, one can easily deduce that the edge mode volume increases with decreasing strip width. At $w$=\SI{100}{\nano\meter}, the edge mode significantly extends over the strip cross section, as expected, due to the increased exchange interaction.

An intricate plethora of the large-$n$ standing spin-wave modes with closely spaced resonant fields/frequencies is observed in relatively wide strips (e.g. $w=$~\SI{500}{nm} and above) for fields around and above the edge saturation field. For reduced strip width, the number of the modes the strip pair can host decreases, and for $w$ below $\sim$~\SI{200}{nm}, only few standing spin waves are present [see Fig.~\figref{fig4}{c,d}]. Note, for the pair of 100-nm-wide strips only three modes can be efficiently excited with spatially homogeneous rf-field at high static fields [see Fig.~\figref{fig4}{d,h}], i.e. the two true edge modes ($n$=1 and $n$=1i) localized at the outer and inner edges (with the closely spaced frequencies) and the quasi-uniform mode localized at the strip center. Furthermore, for such narrow strips only the edge mode softening remains, showing that the magnetization of the strip across the width rotates almost homogeneously into the direction of the static field. At about \SI{360}{\milli\tesla} the frequency drops to a minimum, close to zero. This field now is related to the shape anisotropy of the strip pair.

In conclusion, we studied the spatial distribution of the standing spin-wave modes in infinitely long Permalloy strips with rectangular cross-section. We define the resonance conditions (fields and frequencies) for the excitation of the true edge modes localized at both inner and outer edges of the neighbouring strips. Furthermore, we study the mode localization in pairs of Permalloy strips as a function of the strip width and the lateral separation between the strips. We show that a wide-range-tunability of the localized edge mode resonances can be achieved with a precise control of the magnetostatic coupling between the strips. The observed mode confinement allows for an efficient control of the SW localization in the pair of sub-100-nm waveguides, with the field-controllable switching between the propagation channels localized either in the strip center or at the edges of the strips [see Fig.~\figref{fig4}{h}]. This study provides a micromagnetic background for the understanding of the interaction between the standing (as well as propagating) spin waves in closely packed magnonic waveguides.

\begin{acknowledgements}
Financial support by the Deutsche Forschungsgemeinschaft (DFG) within the programs IU 5/2-1 (project number 501377640) and KA 5069/3-1 (project number 444929866) is gratefully acknowledged.
\end{acknowledgements}

\bibliographystyle{apsrev4-2}
\bibliography{references.bib}

\end{document}